\documentclass[a4paper,10pt,twoside]{cpc-hepnp}
\usepackage{mathrsfs}
\usepackage{epsfig}
\usepackage{textcomp}
\usepackage{amsmath}
\usepackage{graphicx}
\usepackage{amssymb}
\usepackage{epstopdf}
\usepackage{slashed}
\usepackage{mathtools}
\usepackage{color}
\usepackage{slashed}
\usepackage{setspace}
\usepackage{subfigure}
\usepackage{graphicx}
\usepackage{float}
\usepackage{caption}
\usepackage{multicol}
\usepackage[colorlinks,linkcolor=blue]{hyperref}
\allowdisplaybreaks[3]

\begin{document}

\title{Categorization of two-loop Feynman diagrams in the $\mathcal O(\alpha^2)$ correction to $e^+e^- \rightarrow ZH$
	\thanks{This work was supported by the National Natural Science Foundation of China under Grant No. 11675185 and 12075251.}
}
\author{
	Zhao Li $^{1,2,3)}$ \email{zhaoli@ihep.ac.cn} \quad
	Yefan Wang$^{1,2)}$ \email{wangyefan@ihep.ac.cn} \quad
	Quan-feng Wu $^{1,2)}$ \email{wuquanfeng@ihep.ac.cn}
}

\maketitle
\address {$^1$ Institute of High Energy Physics, Chinese Academy of Sciences, Beijing 100049, China \\
    $^2$ School of Physics Sciences, University of Chinese Academy of Sciences, Beijing 100039, China \\
    $^3$ Center of High Energy Physics, Peking University, Beijing 100871, China}

\begin{abstract}
The $e^+e^- \rightarrow ZH$ process is the dominant process for the Higgs boson production at the future Higgs factory. 
In order to match the analysis on the Higgs properties with the highly precise experiment data, 
it will be crucial to include the theoretical prediction to the full next-to-next-to-leading order electroweak effect in 
the production rate $\sigma(e^+e^-\rightarrow ZH)$.
In this inspiring work, we categorize the two-loop Feynman diagrams of the $\mathcal O(\alpha^2)$ correction to $e^+e^- \rightarrow ZH$ 
into 6 categories according to the relevant topological structures. 
Although 25377 diagrams contribute to the $\mathcal O(\alpha^2)$ correction in total, 
the number of the most challenging diagrams with seven denominators is 2250, 
which contain only 312 non-planar diagrams with 155 independent types. 
This categorization could be a valuable reference for the complete calculation in the future.
\end{abstract}


\begin{multicols}{2}
\section{Introduction}
The discovery of the Higgs boson in 2012 at the Large Hadron Collider (LHC) \cite{Aad:2012tfa,Chatrchyan:2012ufa} has opened a new era in the particle physics. In particular, the Higgs boson is regarded as the key to solve some challenging problems such as the problem of hierarchy, the origin of the neutrino mass and the dark matter problem. Consequently the precise measurements of the Higgs boson properties have become top priorities of both experimental and theoretical particle physics. 

Although the LHC can produce a lot of Higgs bosons, the enormously complicated QCD backgrounds make the sufficiently precise measurements hard to achieve. 
To precisely measure the properties of the Higgs boson, the next generation $e^+e^-$ colliders have been proposed as the Higgs factories aiming at much higher accuracy of measurements. Compared to the LHC, the $e^+e^-$ colliders will have cleaner experiment conditions and higher luminosity. The candidates of the next generation $e^+e^-$ colliders include Circular Electron Positron Collider (CEPC) \cite{CEPCStudyGroup:2018ghi,CEPCStudyGroup:2018rmc}, International Linear Collider (ILC) \cite{Baer:2013cma,Behnke:2013xla,Bambade:2019fyw} and Future Circular Collider (FCC-ee) \cite{Gomez-Ceballos:2013zzn,Abada:2019zxq,Abada:2019lih}. All of them are designed to operate at the center-of-mass energy $\sqrt{s} \sim 240 - 250$ GeV. In this energy range, the processes to produce Higgs bosons are $e^+e^- \rightarrow ZH$ (Higgsstrahlung), $e^+e^- \rightarrow \nu_e \bar \nu_e H$ ($W$ fusion) and $e^+e^- \rightarrow e^+e^- H$ ($Z$ fusion). The dominant Higgs production process is the Higgsstrahlung. And the recoil mass method can be applied to identify the Higgs boson candidates \cite{Ioffe:1976sd,McCullough:2013rea,CEPCStudyGroup:2018ghi,Yan:2016xyx,Abada:2019zxq}. Then the Higgs boson production can be disentangled in a model-independent way. 

At the CEPC, over one million Higgs bosons can be produced in total with the expected integrated luminosity of $5.6 \text{ ab}^{-1}$ \cite{An:2018dwb}. With these sizable events, many important properties of the Higgs boson can be measured in a high accuracy. For instance, the cross section $\sigma(e^+e^-\rightarrow ZH)$ can be measured to the extremely high precision $0.51\%$ \cite{An:2018dwb}. Since the Higgs boson candidate events can be identified by recoil mass method, the measurements of $HZZ$ coupling mainly depend on the precise measurement of $\sigma(e^+e^-\rightarrow ZH)$. Consequently, it is expected that the experiment error in the $HZZ$ coupling can be $0.25\%$ at the CEPC, which is much better than at the HL-LHC \cite{ATLAS:2018jlh,CMS:2013xfa}. 

Furthermore, such precise measurements give the CEPC unprecedented reach into the new physics scenarios which are difficult to probe at the LHC \cite{Craig:2014una}. In the natural supersymmetry (SUSY), typically dominant effect on Higgs precision from colored top partners may have blind spots \cite{Ellwanger:2009dp,Fan:2014axa}. The blind spots can be filled in by the measurement of $\sigma(e^+e^-\rightarrow ZH)$, which is sensitive to loop-level corrections to the tree-level $HZZ$ coupling \cite{Craig:2014una}. When the $\delta \sigma_{ZH}$ approaches to $0.2\%$, the further constraints for $m_{\tilde t_1}$ and $m_{\tilde t_2}$ can be observed \cite{Essig:2017zwe}. And the $HZZ$ coupling plays an important role in the study of Electroweak Phase Transition (EWPT). In the real scalar singlet model \cite{Choi:1993cv,McDonald:1993ey,Profumo:2007wc}, the first order phase transition tends to predict a large suppression of the $HZZ$ coupling ranging from $1\%$ to as much as $30\%$ \cite{Huang:2016cjm}. With the expected sensitivity of $\delta g_{HZZ}$ at the CEPC, the models with a strongly first order phase transition can be tested. 

Beside the improvement on the experiment accuracy, the higher precision theoretical prediction for $\sigma(e^+e^-\rightarrow ZH)$ is also demanded to match the precision of experiment measurements.
The next-to-leading-order (NLO) electroweak (EW) corrections to $\sigma(e^+e^-\rightarrow ZH)$ have been investigated two decades ago \cite{Fleischer:1982af,Kniehl:1991hk,Denner:1992bc}. And the next-to-next-to-leading-order (NNLO) EW-QCD corrections have also been calculated in recent years \cite{Gong:2016jys,Sun:2016bel,Chen:2018xau}. The results show that the NNLO EW-QCD corrections increase the cross section more than one percent, which is larger than the expected experiment accuracies of the CEPC. On the other hand, it indicates the NNLO EW corrections can be significant. And it is necessary to emphasize that the corrections depend crucially on the renormalization schemes. In $\alpha(0)$ scheme, the NNLO EW-QCD corrections are about $1.1\%$ of the leading-order (LO) cross section. But in $G_\mu$ scheme, the NNLO EW-QCD corrections only amount to $0.3\%$ of LO cross section \cite{Sun:2016bel}. And the sensitivity to the different scheme is reduced by NNLO EW-QCD corrections compared to NLO EW corrections. Consequently the EW-QCD $\sigma_{\text{NNLO}}$ ranges from $231\text{ fb}$ to $233\text{ fb}$.

Hence, the missing two-loop corrections to $\sigma(e^+e^-\rightarrow ZH)$ can still lead to an intrinsic uncertainty of $\mathcal O (1\%)$ \cite{Freitas:2019bre}, which is still larger than the experiment accuracy. Since $\sigma(e^+e^-\rightarrow ZH)$ is proportional to the square of the $HZZ$ coupling, the theoretical uncertainties also have a significant impact on the extracting of $HZZ$ coupling. Then the accuracy of $HZZ$ coupling ($0.25\%$) in the CEPC may not be achieved due to the large theoretical uncertainties. Recently, some interesting calculations toward NNLO EW correction have been made such as \cite{Song:2021vru}. We believe that if the full NNLO EW corrections to $\sigma(e^+e^-\rightarrow ZH)$ can be calculated, the scheme dependence can be further reduced to stabilize the theoretical prediction. 

Due to the complicacy of EW interaction, there are more than 20 thousand Feynman diagrams contribute to the $\mathcal O(\alpha^2)$ correction of $e^+e^- \rightarrow ZH$. The complete calculations of these Feynman diagrams are huge projects. Therefore, in this paper we focus on the categorization of these Feynman diagrams. This categorization could be helpful for the future calculations and analyses. In the next section we will categorize the Feynman diagrams into six categories and dozens of subcategories. Then the conclusion is made. 

\section{Categorization}
In the Feynman gauge, we obtain 25377 diagrams contributing to the $\mathcal O(\alpha^2)$ corrections of $e^+e^- \rightarrow ZH$ by using \textsc{FeAmGen}\footnote{The Julia package (\url{https://github.com/zhaoli-IHEP/FeAmGen.jl}) interfaced to \textsc{Qgraf} \cite{Nogueira:1991ex}.}. Here we have chosen the Yukawa couplings of light fermions (all fermions except the top quark) as zero. \textsc{FeynArts} \cite{Hahn:2000kx} is used to check the correctness to this procedure.

To categorize these diagrams, first we put the diagrams which can be factorized into two one-loop diagrams to the category $\mathcal C_1$. Then according to the number of denominators in each diagram, we categorize the remaining non-factorizable two-loop diagrams into five categories $\mathcal C_2,\dots,\mathcal C_6$. Furthermore, according to the topologies of loop structures $\mathcal C_i$ can be categorized into several subcategories $\{\mathcal C_{i,j}\}$. Since the light quarks are regarded as massless except the top quark, we use $\mathcal C_{i,j,a}$ ($\mathcal C_{i,j,b}$) to denote the diagrams without (with) the top quark.\footnote{The complete PDF files of all subcategories can be downloaded in \url{https://github.com/zhaoli-IHEP/eeHZ\_nnloEW\_diagrams}.}.
Since some amplitudes can be obtained by replacing coupling factors or masses from other diagrams, $\mathcal C_{i,j}$ can be reduced to subset $\mathcal C_{i,j}^{ind}$ which only includes the "independent" diagrams. Due to the color structure and conservation laws, 153 diagrams in total have zero amplitudes.

\begin{figure}[H]
	\begin{center}
        \subfigure[]{
            \centering
            \includegraphics[width=2.6cm]{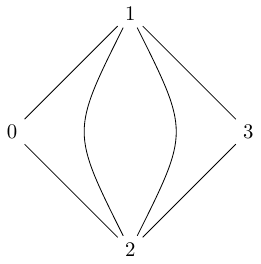}
            \label{fig:nickel-a}
        }\hspace{3mm}
        \subfigure[]{
            \centering
            \includegraphics[height=2.6cm]{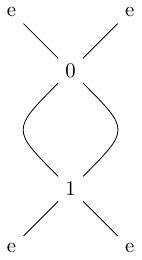}
            \label{fig:nickel-b}
        }
    \caption{Nickel notation and Nickel index}
    \end{center}
\end{figure}

\begin{figure}[H]
	\begin{center}
	\subfigure[]{
		\centering
		\includegraphics[width=4cm]{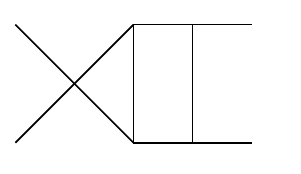}	
	}\hspace{-4mm}
	\subfigure[]{
		\centering
		\includegraphics[width=4cm]{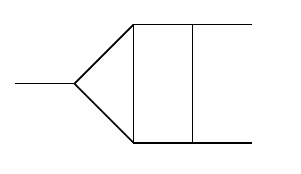}	
    }
    \caption{The example of equivalent topological structures}
    \label{fig:equiv-topo}
\end{center}
\end{figure}

In this paper we use the Nickel index \cite{Batkovich:2014bla,nickel,Nagle:1966} to describe the topologies of loop structures. For reader's convenience we briefly explain the Nickel notation and Nickel index.
Nickel notation is a labelling algorithm to describe connected undirected graphs with "simple" edges and vertices such as the topological structures of the Feynman diagrams. First, one should consider a connected graph with $n$ vertices and label these $n$ vertices by the integers $0$ through $n-1$ at random. Therefore, the sequence can be constructed according to \cite{nickel, Batkovich:2014bla}:
\begin{equation}\begin{aligned}
	& \text{vertices connected to vertex 0 } | \\
	& \text{vertices connected to 1 excluding 0 } |\ \cdots\ |\\
	& \text{vertices connected to vertex } i \text{
	excluding 0 through } i-1\ | \\
	& \cdots\ |.
\end{aligned}\end{equation}
For instance, Fig.\ref{fig:nickel-a} can be represented by $12|223|3|$. Otherwise, we can use the label "$e$"s to describe the external lines in the diagrams. Also, the Nickel notation of Fig.\ref{fig:nickel-b} is $ee11|ee|$. With different labeling strategy, one diagram can be represented as some different Nickel notations, which describe the same diagram up to a topological homeomorphism. For simplicity, the Nickel index algorithm can be used to find the "minimal" Nickel notation, which is called Nickel index. Consequently, the diagram and its Nickel index are in one-to-one correspondence. For instance, the Nickel index of Fig.\ref{fig:nickel-a} is $1123|23|||$. The package \textsc{GraphState} \cite{Batkovich:2014bla} is the useful tool for constructing the Nickel index. The detail of the Nickel index algorithm can be found in Ref. \cite{Batkovich:2014bla}.
In this paper, the topological structures of the diagrams with one vertex connecting to one or two external legs are regarded as equivalent. For instance, the topological structures of two diagrams in Fig.\ref{fig:equiv-topo} can be regarded as equivalent.

\subsection{Category $\mathcal C_1$}
The category $ \mathcal C_1$ includes 7908 Feynman diagrams that can be factorized into two one-loop diagrams. Therefore, the calculations of diagrams in $ \mathcal C_1$ can be regarded as the one-loop level calculations. According to the topologies of loop structures in $ \mathcal C_1$, they are categorized into 3 subcategories. 

The subcategory $\mathcal C_{1,1}$ includes 2117 diagrams which contain at least one one-loop vacuum bubble diagram. Furthermore $\mathcal C_{1,1,a}$ includes 2055 diagrams and $\mathcal C_{1,1,b}$ includes 62 diagrams. In $\mathcal C_{1,1,b}$, there are 14 diagrams whose amplitudes equal to zero.
$\mathcal C_{1,1,a}^{ind}$ has 449 independent diagrams and $\mathcal C_{1,1,b}^{ind}$ has 36 independent diagrams.
We choose diagram \#47 as the representative of $\mathcal C_{1,1,a}$ and diagram \#4418 as the representative of $\mathcal C_{1,1,b}$. 
\begin{figure}[H]
    \centering
    \includegraphics[width=0.4\textwidth]{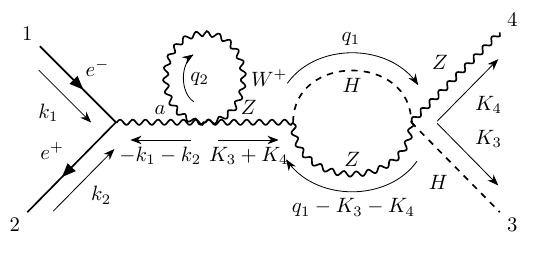}
    \caption{Diagram \#47 (representative of $\mathcal C_{1,1,a}$) }
\end{figure}
\begin{figure}[H]
	\centering
	\includegraphics[width=0.45\textwidth]{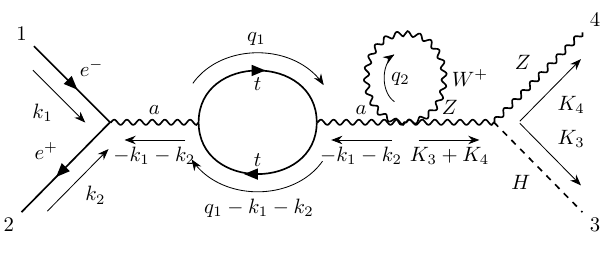}
	\caption{Diagram \#4418 (representative of $\mathcal C_{1,1,b}$) }
\end{figure}
The subcategory $\mathcal C_{1,2}$ includes 5513 diagrams that contain self-energy corrections. And the diagrams $\mathcal C_{1,2}$ do not contain vacuum bubble diagrams. $\mathcal C_{1,2,a}$ includes 4775 diagrams and $\mathcal C_{1,2,b}$ includes 738 diagrams. In $\mathcal C_{1,2,b}$ there are 131 diagrams whose amplitudes equal to zero.
$\mathcal C_{1,2,a}^{ind}$ has 740 independent diagrams and $\mathcal C_{1,2,b}^{ind}$ has 278 independent diagrams.
We choose diagram \#36 as the representative of $\mathcal C_{1,2,a}$ and diagram \#1035 as the representative of $\mathcal C_{1,2,b}$. 
\begin{figure}[H]
	\centering
	\includegraphics[width=0.45\textwidth]{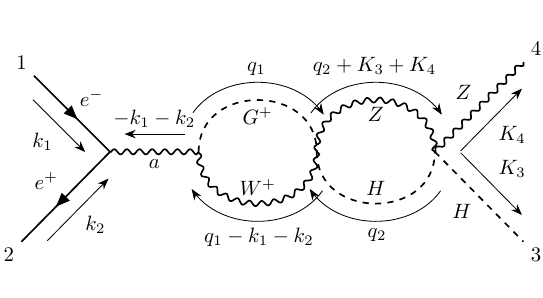}
	\caption{Diagram \#36 (representative of $\mathcal C_{1,2,a}$) }
\end{figure}
\begin{figure}[H]
	\centering
	\includegraphics[width=0.4\textwidth]{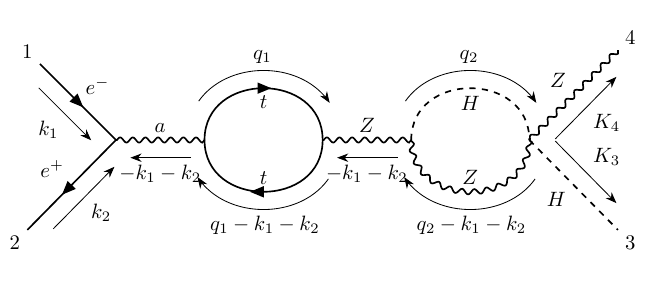}
	\caption{Diagram \#1035 (representative of $\mathcal C_{1,2,b}$) }
\end{figure}
The subcategory $\mathcal C_{1,3}$ includes 278 diagrams which contain two vertex corrections. $\mathcal C_{1,3,a}$ includes 260 diagrams and $\mathcal C_{1,3,b}$ includes 18 diagrams. 
$\mathcal C_{1,3,a}^{ind}$ has 68 independent diagrams and $\mathcal C_{1,3,b}^{ind}$ has 14 independent diagrams.
We choose diagram \#6983 as the representative of $\mathcal C_{1,3,a}$ and diagram \#23660 as the representative of $\mathcal C_{1,3,b}$.
\begin{figure}[H]
	\centering
	\includegraphics[width=0.4\textwidth]{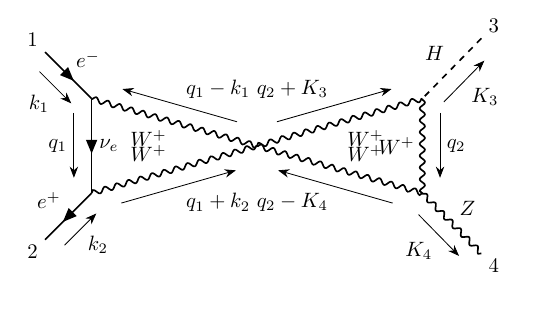}
	\caption{Diagram \#6983 (representative of $\mathcal C_{1,3,a}$) }
\end{figure}
\begin{figure}[H]
	\centering
	\includegraphics[width=0.4\textwidth]{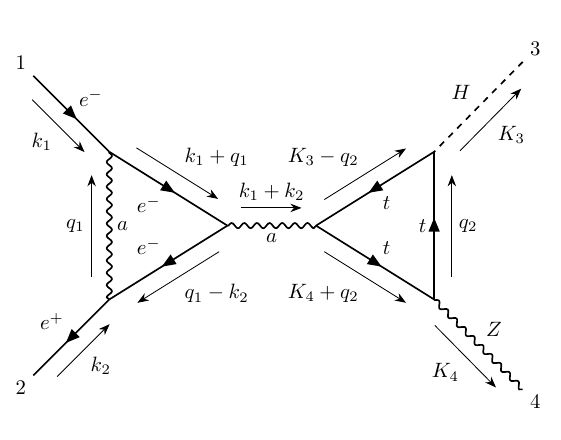}
	\caption{Diagram \#23660 (representative of $\mathcal C_{1,3,b}$) }
\end{figure}

\subsection{Category $\mathcal C_2$}
The category $ \mathcal C_2$ includes non-factorizable two-loop Feynman diagrams with three denominators. We found that all diagrams in $\mathcal C_{2}$ are two-loop self-energy diagrams. $\mathcal C_2$ includes 18 diagrams, none of which contains top quark. 
$\mathcal C_{2}^{ind}$ has 8 independent diagrams.
We choose diagram \#519 as the representative of $ \mathcal C_2$.
\begin{figure}[H]
    \centering
    \includegraphics[width=0.4\textwidth]{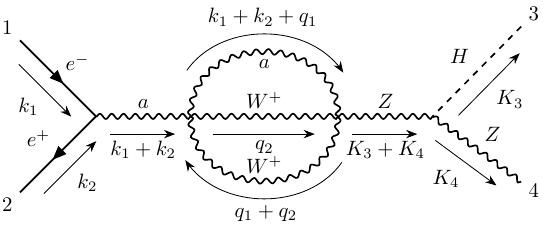}
    \caption{Diagram \#519 (representative of $\mathcal C_{2}$)}
\end{figure}

\subsection{Category $\mathcal C_3$}
The category $ \mathcal C_3$ includes 593 non-factorizable two-loop Feynman diagrams with four denominators. According to the topologies of loop structures in $ \mathcal C_3$, we categorize them into 3 subcategories. 

The subcategory $\mathcal C_{3,1}$ includes 142 diagrams which can be separated into the tree-level diagrams and the two-loop vacuum bubble diagrams. The topology of their loop structures can be noted as $112|2||$ in Nickel index. The calculation of the two-loop vacuum bubble diagram has been well studied \cite{Davydychev:1992mt}.  
$\mathcal C_{3,1}^{ind}$ has 51 independent diagrams.
We choose diagram \#3961 as the representative of $ \mathcal C_{3,1}$.
\begin{figure}[H]
    \centering
    \includegraphics[width=0.4\textwidth]{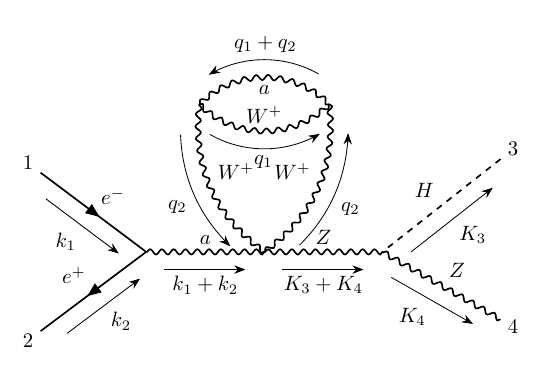}
    \caption{Diagram \#3961 (representative of $\mathcal C_{3,1}$)}
\end{figure}
The subcategory $\mathcal C_{3,2}$ includes 337 two-loop self-energy diagrams, none of which contains top quark. 
$\mathcal C_{3,2}^{ind}$ has 93 independent diagrams.
We choose diagram \#1 as the representative of $ \mathcal C_{3,2}$.
\begin{figure}[H]
    \centering
    \includegraphics[width=0.4\textwidth]{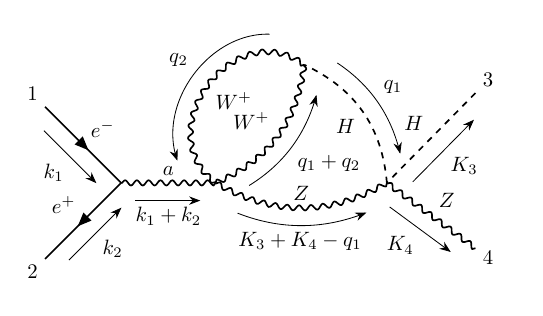}
    \caption{Diagram \#1 (representative of $\mathcal C_{3,2}$)}
\end{figure}
The subcategory $\mathcal C_{3,3}$ includes 114 two-loop vertex correction diagrams, none of which contains top quark. The topology of their loop structures can be noted as $e112|e2|e|$ in Nickel index. And the denominators of diagrams in $\mathcal C_{3,3}$ only depend on two external momenta.  
$\mathcal C_{3,3}^{ind}$ has 24 independent diagrams.
We choose diagram \#191 as the representative of $ \mathcal C_{3,3}$.
\begin{figure}[H]
    \centering
    \includegraphics[width=0.4\textwidth]{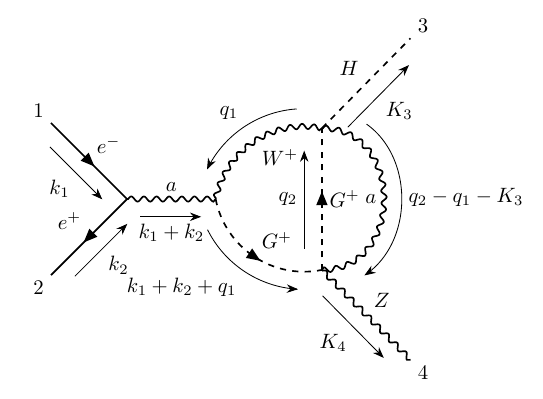}
    \caption{Diagram \#191 (representative of $\mathcal C_{3,3}$)}
\end{figure}

\subsection{Category $\mathcal C_4$}
The category $ \mathcal C_4$ includes 4773 non-factorizable two-loop Feynman diagrams with five denominators. According to the topologies of loop structures in $ \mathcal C_4$, we categorize them into 3 subcategories. 

The subcategory $\mathcal C_{4,1}$ includes 3266 two-loop self-energy diagrams, some of which contain top quark. $\mathcal C_{4,1,a}$ includes 2565 diagrams and $\mathcal C_{4,1,b}$ includes 701 diagrams. 
$\mathcal C_{4,1,a}^{ind}$ has 753 independent diagrams and $\mathcal C_{4,1,b}^{ind}$ has 249 independent diagrams.
We choose diagram \#603 as the representative of $\mathcal C_{4,1,a}$ and diagram \#611 as the representative of $\mathcal C_{4,1,b}$. 
\begin{figure}[H]
    \centering
    \includegraphics[width=0.4\textwidth]{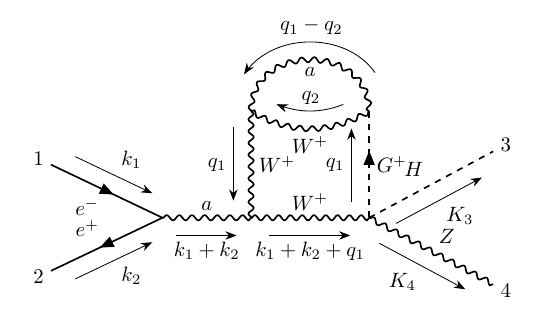}
    \caption{Diagram \#603 (representative of $\mathcal C_{4,1,a}$)}
\end{figure}
\begin{figure}[H]
    \centering
    \includegraphics[width=0.4\textwidth]{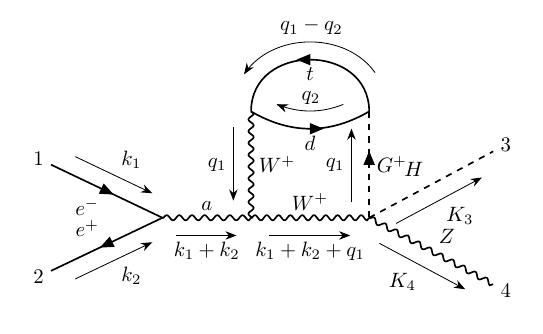}
    \caption{Diagram \#611 (representative of $\mathcal C_{4,1,b}$)}
\end{figure}
The subcategory $\mathcal C_{4,2}$ includes 637 two-loop vertex correction diagrams, none of which contains top quark. The topology of their loop structures can be noted as $e12|e23|3|e|$ in Nickel index. And the denominators of diagrams in $\mathcal C_{4,2}$ only depend on two external momenta.
$\mathcal C_{4,2}^{ind}$ has 140 independent diagrams.
 We choose diagram \#2676 as the representative of $\mathcal C_{4,2}$. 
\begin{figure}[H]
    \centering
    \includegraphics[width=0.4\textwidth]{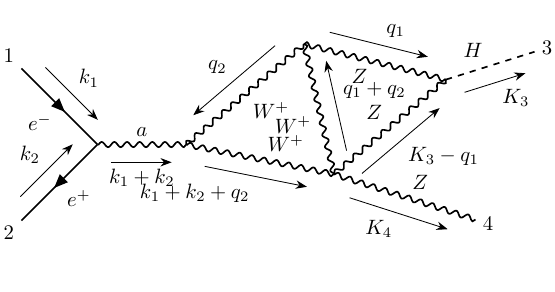}
    \caption{Diagram \#2676 (representative of $\mathcal C_{4,2}$)}
\end{figure}
The subcategory $\mathcal C_{4,3}$ includes 870 two-loop vertex correction diagrams, none of which contains top quark. The topology of their loop structures can be noted as $e112|3|e3|e|$ in Nickel index. And the denominators of diagrams in $\mathcal C_{4,3}$ only depend on two external momenta. 
$\mathcal C_{4,3}^{ind}$ has 278 independent diagrams.
We choose diagram \#3063 as the representative of $\mathcal C_{4,3}$.
\begin{figure}[H]
    \centering
    \includegraphics[width=0.4\textwidth]{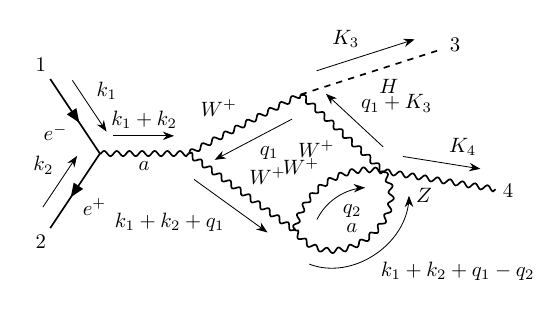}
    \caption{Diagram \#3063 (representative of $\mathcal C_{4,3}$)}
\end{figure}

\subsection{Category $ \mathcal C_5$}
The category $ \mathcal C_5$ includes 9835 non-factorizable two-loop Feynman diagrams with six denominators. According to the topologies of loop structures in $ \mathcal C_5$, we categorize them into six subcategories.

The subcategory $\mathcal C_{5,1}$ includes two-loop planar triangle diagrams. The topology of their loop structures can be noted as $e12|e3|34|4|e|$ in Nickel index. And the denominators of diagrams in $\mathcal C_{5,1}$ only depend on two external momenta. $\mathcal C_{5,1}$ includes 4897 diagrams, some of which contain top quark. Then $\mathcal C_{5,1,a}$ includes 3966 diagrams and $\mathcal C_{5,1,b}$ includes 931 diagrams. 
$\mathcal C_{5,1,a}^{ind}$ has 1039 independent diagrams and $\mathcal C_{5,1,b}^{ind}$ has 397 independent diagrams.
We choose diagram \#1325 as the representative of $\mathcal C_{5,1,a}$ and diagram \#16206 as the representative of $\mathcal C_{5,1,b}$. 
\begin{figure}[H]
    \centering
    \includegraphics[width=0.4\textwidth]{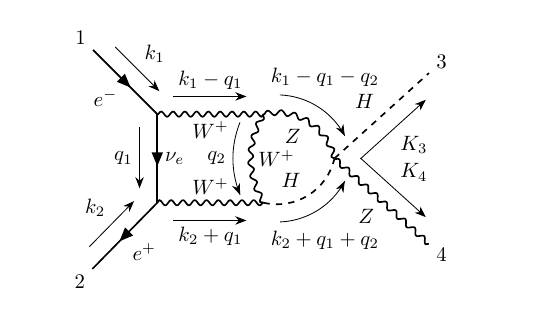}
    \caption{Diagram \#1325 (representative of $\mathcal C_{5,1,a}$)}
\end{figure}
\begin{figure}[H]
    \centering
    \includegraphics[width=0.4\textwidth]{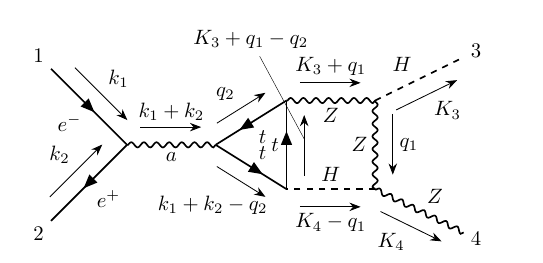}
    \caption{Diagram \#16206 (representative of $\mathcal C_{5,1,b}$)}
\end{figure}
The subcategory $\mathcal C_{5,2}$ includes 184 two-loop planar diagrams, none of which contains top quark. The topology of their loop structures can be noted as $e12|e23|4|e4|e|$ in Nickel index.  
$\mathcal C_{5,2}^{ind}$ has 90 independent diagrams.
We choose diagram \#3613 as the representative of $ \mathcal C_{5,2}$.
\begin{figure}[H]
    \centering
    \includegraphics[width=0.4\textwidth]{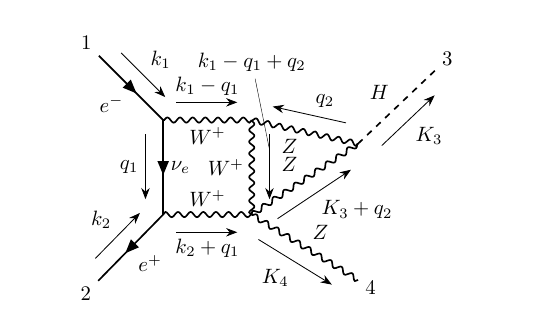}
    \caption{Diagram \#3613 (representative of $\mathcal C_{5,2}$)}
\end{figure}
The subcategory $\mathcal C_{5,3}$ includes two-loop planar diagrams. The topology of their loop structures can be noted as $e12|e3|e4|44||$ in Nickel index. And the denominators of diagrams in $\mathcal C_{5,3}$ only depend on two external momenta. $\mathcal C_{5,3}$ includes 4067 diagrams, some of which contains top quark. $\mathcal C_{5,3,a}$ includes 3260 diagrams and $\mathcal C_{5,3,b}$ includes 807 diagrams. In $\mathcal C_{5,3,b}$, there are 131 diagrams whose amplitudes equal to zero. $\mathcal C_{5,3,a}^{ind}$ has 1077 independent diagrams and $\mathcal C_{5,3,b}^{ind}$ has 264 independent diagrams.
We choose diagram \#14794 as the representative of $\mathcal C_{5,3,a}$ and diagram \#14812 as the representative of $\mathcal C_{5,3,b}$. 
\begin{figure}[H]
    \centering
    \includegraphics[width=0.4\textwidth]{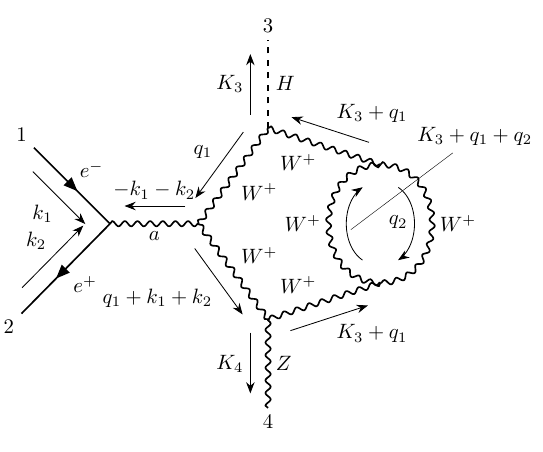}
    \caption{Diagram \#14794 (representative of $\mathcal C_{5,3,a}$)}
\end{figure}
\begin{figure}[H]
    \centering
    \includegraphics[width=0.4\textwidth]{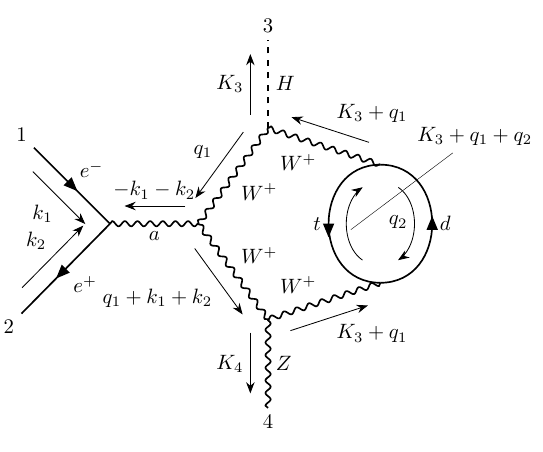}
    \caption{Diagram \#14812 (representative of $\mathcal C_{5,3,b}$)}
\end{figure}
The subcategory $\mathcal C_{5,4}$ includes two-loop planar diagrams. The topology of their loop structures can be noted as $e112|3|e4|e4|e|$ in Nickel index. $\mathcal C_{5,4}$ includes 116 Feynman diagrams, none of which contains top quark.
$\mathcal C_{5,4}^{ind}$ has 70 independent diagrams.
We choose diagram \#3845 as a representative of $\mathcal C_{5,4}$.
\begin{figure}[H]
    \centering
    \includegraphics[width=0.4\textwidth]{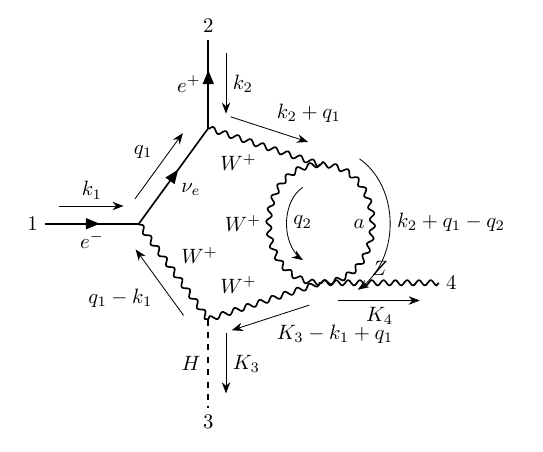}
    \caption{Diagram \#3845 (representative of $\mathcal C_{5,4}$)}
\end{figure}
The subcategory $\mathcal C_{5,5}$ includes two-loop non-planar triangle diagrams. The topology of their loop structures can be noted as $e12|34|34|e|e|$ in Nickel index. And the denominators of diagrams in $\mathcal C_{5,5}$ only depend on two external momenta. $\mathcal C_{5,5}$ includes 560 diagrams, some of which contain top quark. $\mathcal C_{5,5,a}$ includes 442 diagrams and $\mathcal C_{5,5,b}$ includes 118 diagrams. 
$\mathcal C_{5,5,a}^{ind}$ has 140 independent diagrams and $\mathcal C_{5,5,b}^{ind}$ has 54 independent diagrams.
We choose diagram \#1267 as the representative of $\mathcal C_{5,5,a}$ and diagram \#11100 as the representative of $\mathcal C_{5,5,b}$. 
\begin{figure}[H]
    \centering
    \includegraphics[width=0.4\textwidth]{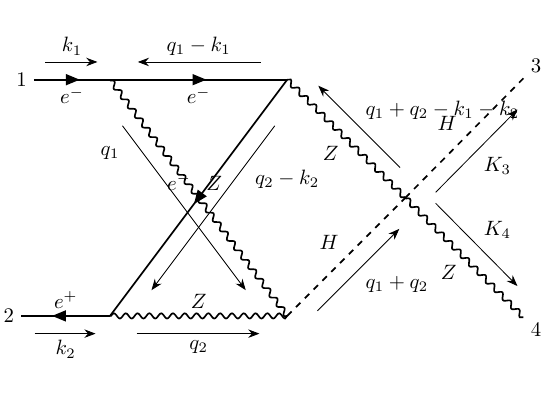}
    \caption{Diagram \#1267 (representative of $\mathcal C_{5,5,a}$)}
\end{figure}
\begin{figure}[H]
    \centering
    \includegraphics[width=0.45\textwidth]{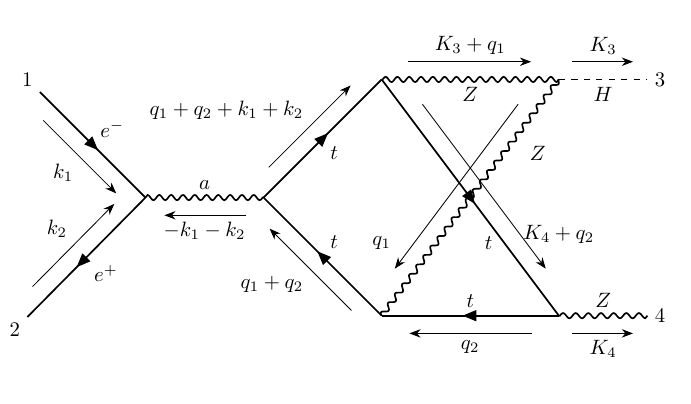}
    \caption{Diagram \#11100 (representative of $\mathcal C_{5,5,b}$)}
\end{figure}
The subcategory $\mathcal C_{5,6}$ includes two-loop non-planar diagrams. The topology of their loop structures can be noted as $e12|e34|34|e|e|$ in Nickel index. $\mathcal C_{5,6}$ includes 11 diagrams, none of which contains top quark. 
$\mathcal C_{5,6}^{ind}$ has 8 independent diagrams.
We choose diagram \#3602 as the representative of $ \mathcal C_{5,6}$.
\begin{figure}[H]
    \centering
    \includegraphics[width=0.4\textwidth]{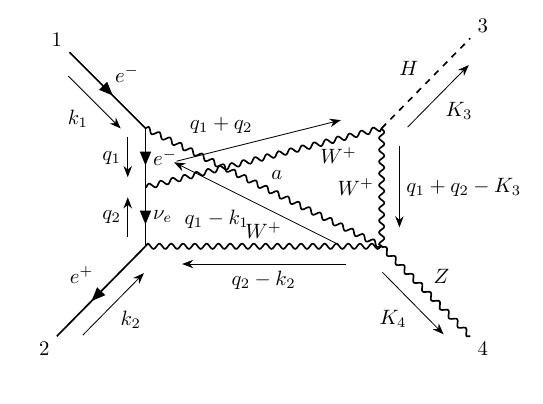}
    \caption{Diagram \#3602 (representative of $\mathcal C_{5,6}$)}
\end{figure}

\subsection{Category $ \mathcal C_6$}
The category $ \mathcal C_6$ includes 2250 non-factorizable two-loop Feynman diagrams with seven denominators. According to the topologies of loop structures in $ \mathcal C_6$, we categorize them into 4 subcategories. 

The subcategory $\mathcal C_{6,1}$ includes 446 two-loop planar double-box diagrams. The topology of their loop structures can be noted as $e12|e3|34|5|e5|e|$ in Nickel index. $\mathcal C_{6,1,a}$ includes 424 diagrams and $\mathcal C_{6,1,b}$ includes 22 diagrams. 
$\mathcal C_{6,1,a}^{ind}$ has 194 independent diagrams and in $\mathcal C_{6,1,b}^{ind}$ has 18 independent diagrams.
We choose diagram \#23202 as the representative of $\mathcal C_{6,1,a}$ and diagram \#23228 as the representative of $\mathcal C_{6,1,b}$.
\begin{figure}[H]
    \centering
    \includegraphics[width=0.4\textwidth]{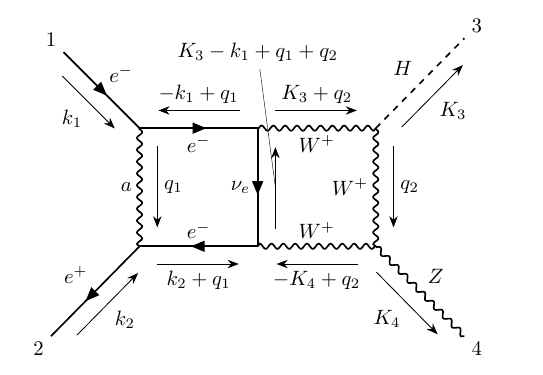}
    \caption{Diagram \#23202 (representative of $\mathcal C_{6,1,a}$)}
\end{figure}
\begin{figure}[H]
    \centering
    \includegraphics[width=0.4\textwidth]{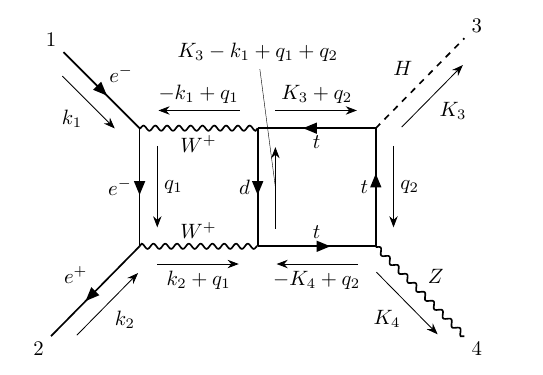}
    \caption{Diagram \#23228 (representative of $\mathcal C_{6,1,b}$)}
\end{figure}
The subcategory $\mathcal C_{6,2}$ includes 688 two-loop planar diagrams. The topology of their loop structures can be noted as $e1|22|3|e4|e5|e6||$ in Nickel index. $\mathcal C_{6,2,a}$ includes 580 diagrams and $\mathcal C_{6,2,b}$ includes 108 diagrams. In $\mathcal C_{6,2,b}$, there are 4 diagrams whose amplitudes equal to zero.
$\mathcal C_{6,2,a}^{ind}$ has 299 independent diagrams and $\mathcal C_{6,2,b}^{ind}$ has 48 independent diagrams.
We choose diagram \#24690 as the representative of $\mathcal C_{6,2,a}$ and diagram \#24708 as the representative of $\mathcal C_{6,2,b}$.
\begin{figure}[H]
    \centering
    \includegraphics[width=0.4\textwidth]{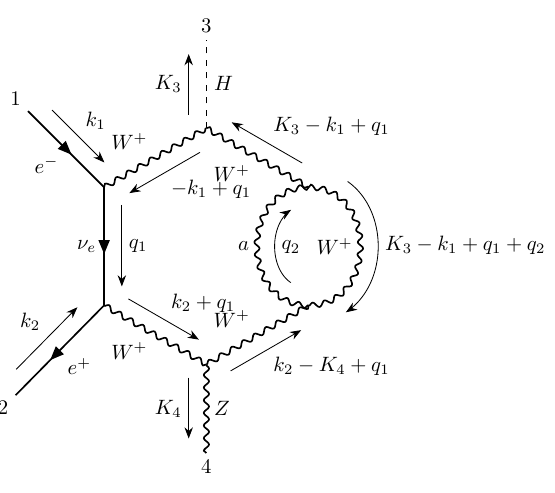}
    \caption{Diagram \#24690 (representative of $\mathcal C_{6,2,a}$)}
\end{figure}
\begin{figure}[H]
    \centering
    \includegraphics[width=0.4\textwidth]{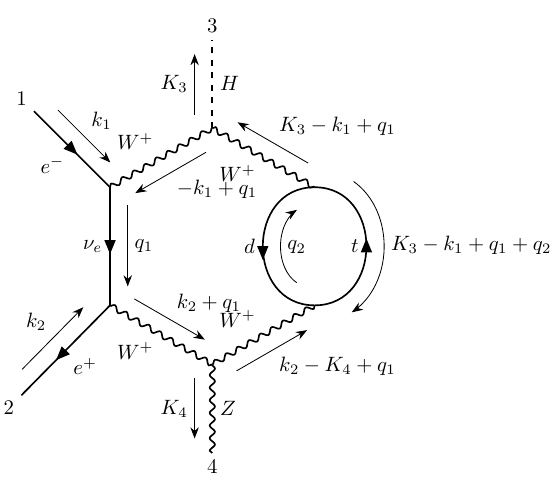}
    \caption{Diagram \#24708 (representative of $\mathcal C_{6,2,b}$)}
\end{figure}
The subcategory $\mathcal C_{6,3}$ includes 804 two-loop planar diagrams. The topology of their loop structures can be represented as $e12|e3|e4|45|5|e|$ in Nickel index. $\mathcal C_{6,3,a}$ includes 733 diagrams and $\mathcal C_{6,2,b}$ includes 71 diagrams. 
$\mathcal C_{6,3,a}^{ind}$ has 302 independent diagrams and $\mathcal C_{6,3,b}^{ind}$ has 42 independent diagrams.
We choose diagram \#23886 as the representative of $\mathcal C_{6,3,a}$ and diagram \#23907 as the representative of $\mathcal C_{6,3,b}$.
\begin{figure}[H]
    \centering
    \includegraphics[width=0.4\textwidth]{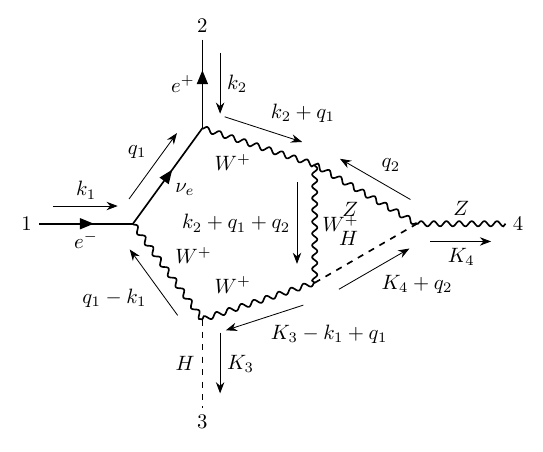}
    \caption{Diagram \#23886 (representative of $\mathcal C_{6,3,a}$)}
\end{figure}
\begin{figure}[H]
    \centering
    \includegraphics[width=0.4\textwidth]{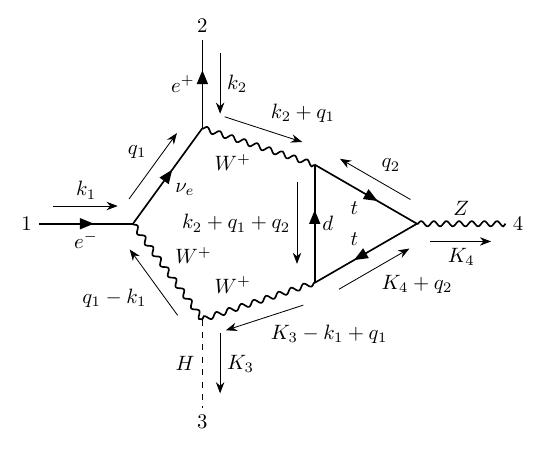}
    \caption{Diagram \#23907 (representative of $\mathcal C_{6,3,b}$)}
\end{figure}
The subcategory $\mathcal C_{6,4}$ is the most challenging subcategory which includes 312 two-loop non-planar double-box diagrams. The topology of their loop structures can be noted as $e12|e3|45|45|e|e|$ in Nickel index. $\mathcal C_{6,4,a}$ includes 301 diagrams and $\mathcal C_{6,4,b}$ includes 11 diagrams.
$\mathcal C_{6,4,a}^{ind}$ has 146 independent diagrams and $\mathcal C_{6,4,b}^{ind}$ has 9 independent diagrams.
We choose diagram \#22890 as the representative of $\mathcal C_{6,4,a}$ and diagram \#22909 as the representative of $\mathcal C_{6,4,b}$.
\begin{figure}[H]
    \centering
    \includegraphics[width=0.4\textwidth]{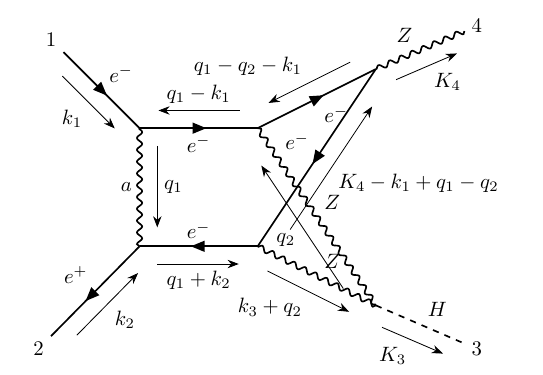}
    \caption{Diagram \#22890 (representative of $\mathcal C_{6,4,a}$)}
\end{figure}
\begin{figure}[H]
    \centering
    \includegraphics[width=0.4\textwidth]{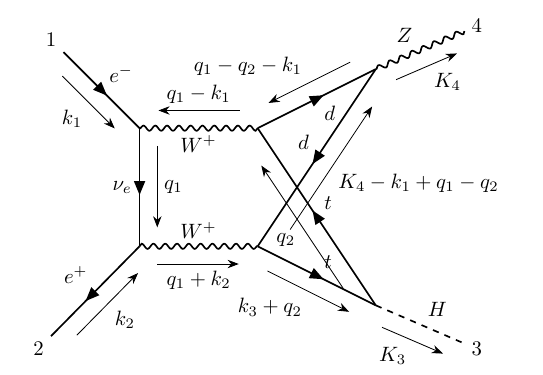}
    \caption{Diagram \#22909 (representative of $\mathcal C_{6,4,b}$)}
\end{figure}
Finally the information for all subcategories has been summarized in Table \ref{table1}.
\end{multicols}
\begin{table}[H]
\centering
\caption{Summary table for all subcategories}
\label{table1}
\begin{tabular}{|c|c|c|c|c|c|}
	\hline
	\begin{tabular}{c}subcategory \\name\end{tabular}& \begin{tabular}{c}number of \\diagrams\end{tabular}&\begin{tabular}{c}number of \\denominators\end{tabular} &   \begin{tabular}{c} non-planar \\ diagrams \end{tabular} &\begin{tabular}{c} contains \\top quark \end{tabular}&\begin{tabular}{c}number of \\independent diagrams\end{tabular}\\
	\hline
	$C_{1,1}$&  2117 & - &No&  Yes& 485  \\
	\hline
	$C_{1,2}$&  5513 & - &No&  Yes& 1018\\
	\hline
	$C_{1,3}$&  278  & - &No&  Yes&  82\\
	\hline
	$C_{2}$&    18   & 3 &No&  No&  8\\
	\hline
	$C_{3,1}$&  142 &  4 &No&  No&  51\\
	\hline
	$C_{3,2}$&  337 &  4 &No&  No&  93\\
	\hline
	$C_{3,3}$&  114 &  4 &No&  No&  24\\
	\hline
	$C_{4,1}$&  3266&  5 &No&  Yes&  1002\\
	\hline
	$C_{4,2}$&  637&   5 &No&  No&  140\\
	\hline
	$C_{4,3}$&  870&   5 &No&  No&  278\\
	\hline
	$C_{5,1}$&  4897&  6 &No&  Yes&  1436\\
	\hline
	$C_{5,2}$&  184&   6 &No&  No&  90\\
	\hline
	$C_{5,3}$&  4067&  6 &No&  Yes&  1341\\
	\hline
	$C_{5,4}$&  116&   6 &No&  No&  70\\
	\hline
	$C_{5,5}$&  560&   6 &Yes&  Yes&  194\\
	\hline
	$C_{5,6}$&  11&    6 &Yes&  No&  8\\
	\hline
	$C_{6,1}$&  446&   7 &No&  Yes&  212\\
	\hline
	$C_{6,2}$&  688&   7 &No&  Yes&  347\\
	\hline
	$C_{6,3}$&  804&   7 &No&  Yes&  344\\
	\hline
	$C_{6,4}$&  312&   7 &Yes&  Yes&  155\\
	\hline
\end{tabular}
\end{table}
\begin{multicols}{2}
\section{Conclusion}
In this paper, we categorize the two-loop Feynman
diagrams contributing to the $\mathcal O(\alpha^2)$ corrections in the Higgsstrahlung $e^+e^- \rightarrow ZH$
into 6 categories and dozens of subcategories. The most challenging subcategory is $\mathcal C_{6,4}$, which includes 312 two-loop non-planar double-box diagrams. And there are only 155 independent diagrams in $\mathcal C_{6,4}$.
We hope that the calculations of these Feynman diagrams can be organized conveniently with the help of this categorization.

\section{Acknowledgments}

\acknowledgments{The authors want to thank Ayres Freitas, Hao Liang, Tao Liu for helpful discussions.}

\end{multicols}

\begin{multicols}{2}

\end{multicols}
\end{document}